\documentclass{article}  
\usepackage{graphicx}

\begin{document}

\begin{titlepage}

\begin{center} 

{\bf Quantum Gravity, CPT symmetry and Entangled States}

\vspace{1cm} 

{ \bf Nick E. Mavromatos}

\vspace{0.5cm} 
King's College London, Department of Physics, \\
              Strand, London WC2R 2LS, U.K.  \\
\vspace{0.5cm} 

{\bf Abstract}

\vspace{0.5cm} 
\end{center} 

 There may unique (``smoking-gun'') signatures of the
 breakdown of CPT symmetry, induced in some models of Quantum Gravity entailing decoherence for quantum matter. Such effects can be observed in entangled states of neutral mesons via modifications of the respective Einstein-Podolsky-Rosen (EPR) correlators (``$\omega$-effect). In the talk I discuss
 experimental signatures and bounds of the $\omega$-effect in $\Phi$-and B-factories, and argue that the effect might be falsifiable at the next generation facilities.

\vspace{2cm}



\vspace{2cm} 
\begin{center} 
\emph{Invited Talk at the EXA2008 \& LEAP 2008 joint Conference, \\
Stefan-Meyer-Institut f\"ur subatomare Physik, Austrian Academy of Sciences,
September 15th-19th 2008, Vienna (Austria) }
\end{center} 

\end{titlepage} 

\section{Introduction: Ways of CPT Violation in Quantum Gravity}
\label{sec:1}

The theory of Quantum Gravity is still elusive and far from any experimental verification.
Nevertheless, in the last decade there have been significant improvements in the precision of terrestrial and astrophysical instrumentation, which resulted in stringent bounds being placed on several models of quantum gravity available in the literature so far. Most of these models predict a breakdown of fundamental symmetries, such as (local) Lorentz invariance and CPT symmetry~\cite{review}.

The sensitivity of various experiments to the so-called Physics at the ``Planck scale'', that is the energy scale at which quantum gravity phenomena are expected to set in, is highly model dependent. For instnace, in the modern version of string theory~\cite{polchinski}, which is one of the most popular and thoroughly worked out theoretical frameworks of Quantum Gravity, the quantum gravity scale may not be the familiar one of Planck energy  $M_P = 1.2 \times 10^{19}$ GeV. The string mass scale, $M_s$, where quantum gravity phenomena take place in the higher-than-four-dimensional space times of string theory, is essentially unconstrained theoretically at present, and may be as low as a few TeV. This prompted the excitement for searching for effects of extra dimensions at the Large Hadron Collider (LHC), which has been recently launched at CERN. However, even if the scale of quantum gravity is as high as $M_P \sim 10^{19} {\rm GeV}$, nevertheless there may be predictions that affect the physics at lower scales, especially in models in which the quantum gravitational interactions behave as a `medium' (environment) in which ordinary matter propagates. The medium idea for quantum gravity is primarily due to J.A. Wheeler~\cite{wheeler}, who visualized Space-time at length scales near the Planck length $\ell_P \sim 10^{-35}$~m as having a ``foamy'' structure, that is containing singular quantum fluctuations of the metric field, with non trivial topologies of microscopic size, such as virtual black holes \emph{etc}.
This issue is linked with the fundamental symmetries breakdown by quantum gravity mentioned earlier, in particular local Lorentz invariance and/or CPT symmetry.

 In this conference there are talks~\cite{lehnert}, which deal with phenomenological models and precision tests of Lorentz invariance. The reader should bear in mind that in a model we may have Lorentz-invariance violating effects, but without any CPT Violation in the Hamiltonian. An example is provided by the so-called non-commutative field theory models. In some of them, one can argue~\cite{carrol} that their low-energy continuum space-time description corresponds to effective field theories of the form encountered in the so-called standard model extension of Kostelecky and collaborators~\cite{lehnert} but with only Lorentz violating higher-dimensional operators, while CPT appears unbroken by the effective lagrangian. On the other hand, if CPT is violated, then both Lorentz- and CPT -symmetry violating effects are present. This seems to be a general consequence of
the axiomatic proof of CPT theorem in flat space time models~\cite{greenberg}, which requires
Lorentz-covariant off-shell correlation functions in a relativistic field theory setting.

In this talk, I will concentrate mainly on the breakdown of the CPT symmetry. In fact, there are two ways by which CPT breakdown is encountered in a quantum gravity model. The first is through the non commutativity of a well-defined quantum mechanical CPT operator (which generates the CPT transformations) with the Hamiltonian of the system under consideration. This is the breakdown of CPT symmetry dealt with in standard Lorentz-violating Extensions of the Standard Model (SME), mentioned above~\cite{lehnert,greenberg,carrol}. In the second way of CPT breaking,
the CPT operator is \emph{ill-defined} as a quantum mechanical operator, but in a \emph{perturbative sense} to be described below. This ill-definition is a consequence of the foamy structure~\cite{wheeler} of space time, whereby the quantum metric fluctuations at Planck scales induce \emph{quantum decoherence} of matter propagating in such backgrounds.
For such cases, the particle field theoretic system is simply an \emph{open quantum mechanical system} interacting with the ``environment'' of quantum gravity. The ill definition of the CPT operator in such cases is of more fundamental nature than the mere non commutativity of this operator with the local effective Hamiltonian of the matter system in Lorentz-symmetry violating SME models. In the cases of quantum-gravity induced decoherence the very concept of a local effective Lagrangian may itself break down. R.~Wald~\cite{wald} has elegantly argued, based on elementary quantum mechanical analysis of open systems,
that the CPT operator cannot exist as a well-defined quantum mechanical operator for systems which exhibit quantum decoherence, that is they are characterised by an evolution of initially pure quantum mechanical states to mixed ones, as the time elapses. This was interpreted as a microscopic time arrow in quantum gravitational media, which induce such decoherence, that is unrelated to CP properties. Hence such ``open'' material systems are characterised by ``\emph{intrinsic CPT violation}'', a terminology we shall use from now in order to describe this particular type of CPT symmetry breakdown.
As a result of the weak nature of quantum gravitational interactions, the ill-definition of the
CPT operator is perturbative in the sense that the anti-particle state still exists, but its properties, as compared to the corresponding particle state, which under normal circumstances would be connected by the action of this operator, are modified. The modifications can be perceived~\cite{sarkar} as a result of the dressing of the (anti-)particle states by perturbative interactions expressing the effects of the medium. In such an approach, the Lorentz symmetry aspect is disentangled from the CPT operator ill-defined nature, in the sense that
Lorentz invariance might not be necessarily violated in such systems (Lorentz-invariant decoherence is known to exist, in the sense of decohered systems with modified Lorentz symmetries, though, to take proper account of the open-system character~\cite{millburn}). 

An interesting question, of experimental interest, concerns the possibility of the observer to prepare decoherent-free subspaces in such quantum-gravity entangled systems. If such a possibility could be realized, then one would have a ``weak form of CPT invariance'' characterising
the system~\cite{wald}, in the sense that
the ill-defined nature of the fundamental CPT operator would not show up in any physical quantities measured in Nature, in particular scattering amplitudes.
Although, theoretically, such a possibility is still not understood, nevertheless the question as to whether there are decoherence-free subspaces in quantum-gravity foam situations can be answered experimentally, at least in principle.

It is the point of this talk to tackle this issue by discussing the effects of  this ``intrinsic CPT violation'' in entangled states of mesons in meson factories. As argued in \cite{bmp}, the perturbatively ill-defined nature of  the CPT operator implies modified Einstein-Podolsky-Rosen (EPR) correlations among the entangled states in meson factories, which are uniquely associated with this effect and can be disentangled experimentally from conventional background effects. We termed this effect $\omega$-effect, a nomenclature I shall follow throughout this talk. My point is to argue that,
although in general there seems to be no single figure of merit for CPT Violation, as this is a highly model-dependent issue, nevertheless, these EPR correlators modifications, if true, may constitute ``smoking-gun'' evidence for this particular type of CPT violation and decoherence in Quantum Gravity.

\section{Experimental Signatures of $\omega$-effect in Kaon ($\Phi$-) Factories}
\label{sec:3}

If CPT is \emph{intrinsically} violated,
in the sense of being not well defined due to decoherence~\cite{wald},
the Neutral mesons $K^0$ and ${\overline K}^0$ should \emph{no
longer} be treated as \emph{identical particles}. As a
consequence~\cite{bmp}, the initial entangled state in $\Phi$
factories $|i>$, after the $\Phi$-meson decay, assumes the form:
{\scriptsize $ |i> = {\cal N} \bigg[ \left(|K_S({\vec
k}),K_L(-{\vec k})>
- |K_L({\vec k}),K_S(-{\vec k})> \right)\nonumber +  \omega \left(|K_S({\vec k}), K_S(-{\vec k})> - |K_L({\vec
k}),K_L(-{\vec k})> \right)  \bigg]$}, where $\omega = |\omega |e^{i\Omega}$ is a complex parameter,
parametrizing the intrinsic CPTV modifications of the EPR
correlations.

\begin{figure}
\centering
  \includegraphics[width=3.5cm]{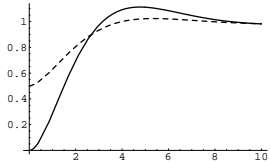}
\caption{A characteristic case of the intensity
$I(\Delta t)$, with $|\omega|=0$ (solid line)  vs  $I(\Delta t)$
(dashed line) with $|\omega|=|\eta_{+-}|$, $\Omega = \phi_{+-} -
0.16\pi$, for definiteness~\cite{bmp}.}
\label{intensomega}
\end{figure}

The $\omega$-parameter controls the amount of contamination of the
final C(odd) state by the ``wrong'' (C(even)) symmetry state.
The appropriate observable (c.f. fig.~\ref{intensomega})
is the ``intensity'' $I(\Delta t)
= \int_{\Delta t \equiv |t_1 - t_2|}^\infty
|A(X,Y)|^2$, with $A(X,Y)$ the appropriate $\Phi$ decay
amplitude~\cite{bmp},
where one of the Kaon products decays to
the  final state $X$ at $t_1$ and the other to the final state $Y$
at time $t_2$ (with $t=0$ the moment of the $\Phi$ decay).
It must be noticed that in Kaon factories there is a particularly good channel,
the one with bi-pion states $\pi^+\pi^-$ as final decay products,
which \emph{enhances the sensitivity}
to the $\omega$-effect by three orders of magnitude. This is due to the fact that
the relevant terms~\cite{bmp} in the intensity $I(\Delta t)$ (c.f. fig.~\ref{intensomega})
contain the combination $\omega/|\eta_{+-}|$, where $\eta_{+-}$ is the relevant CP-violating
amplitude for the $\pi^+\pi^-$ states, which is of order $10^{-3}$.
The KLOE experiment has just released the first measurement of the
$\omega$ parameter~\cite{adidomenico}: $ {\rm Re}(\omega) =
\left( -2.5^{+3.1}_{-2.3}\right)\times 10^{-4}~$, ${\rm
Im}(\omega) = \left( -2.2^{+ 3.4}_{-3.1}\right)\times
10^{-4}$. At least an order of magnitude improvement is expected
for upgraded facilities such as KLOE-2 at (the upgraded)
DA$\Phi$NE-2~\cite{adidomenico}.

This sensitivity is not far from
certain optimistic models of space time foam leading to
$\omega$-like effects~\cite{sarkar}. Indeed, in such models, which are inspired from string theory,
the $\omega$-effect is the result of  local distortions of space time in the neighborhood of space-time
defects, which interact --via topologically non-trivial interactions (string capture/spiltting)--\emph{ only} with electrically neutral matter string states, due to electric charge conservation.
The recoil of the Planck-mass defect results in metric deformations along the direction of motion of the string state, $g_{0i} \sim \Delta k^i /M_P = \zeta k^i /M_P$, where $\Delta k^i = \zeta k^i$ denotes the momentum transfer of the matter state. On average, $\langle \zeta k^i \rangle = 0$, so Lorentz invariance holds macroscopically, but one has non trivial quantum fluctuations $\langle \zeta^2 k_i k_j \rangle \propto \delta_{ij} \overline{\zeta}^2 |\vec k|^2 $. It can be shown~\cite{sarkar} that as a result of such stochastic interactions with the space time foam, neutral entangled states -- such as the ones in meson factories -- exhibit $\omega$-like effects, with the order of magnitude estimate: $|\omega |^2 \sim \frac{|\vec k|^4\overline{\zeta}^2}{M_P^2 (m_1 - m_2)^2}$,
where $m_i$, $i=1,2$ are the masses of the (near degenerate) mass eigenstates. For the energies in an upgrade of DA$\Phi$NE, for instance, it can be seen that $|\omega | \sim 10^{-4}|\overline{\zeta}|$, which lies within the sensitivity of the facility
for values of the average momentum transfer $\overline{\zeta} > 10^{-2}$ (which may be expected on account of naturalness, although this is actually a number that depends on the microscopic quantum theory of gravity under consideration, and hence is still elusive).

We close this section by mentioning that the $\omega$ effect can be
disentangled~\cite{bmp} experimentally from \emph{both}, the C(even) background
- by means of different interference with the C(odd) resonant
contributions - and the decoherent evolution effects of space-time foam~\cite{ehns}, due to different structures.
\section{Experimental Signatures of $\omega$-effect in B-Factories}
\label{sec:4}
\begin{figure}
\centering
  \includegraphics[width=4.5cm]{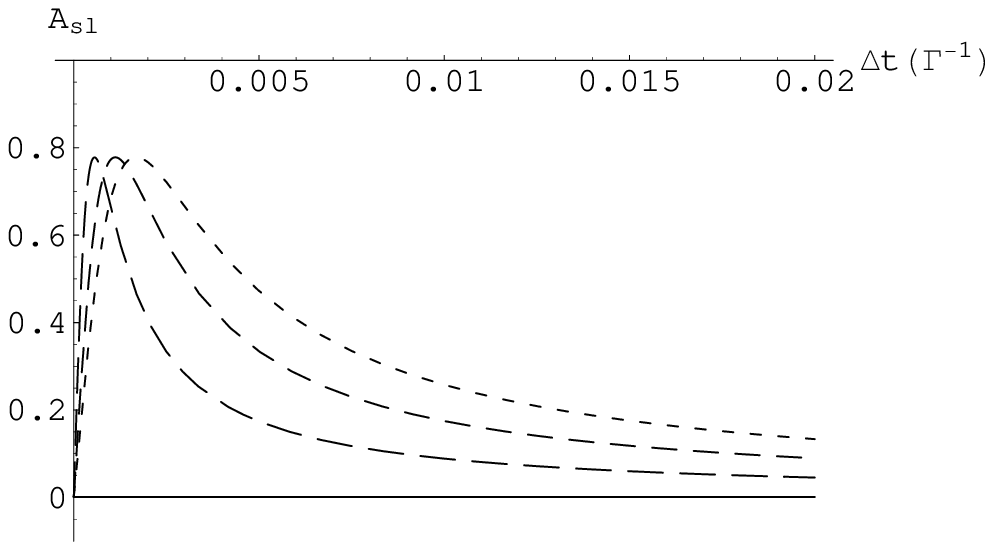}\hfill \includegraphics[width=4.5cm]{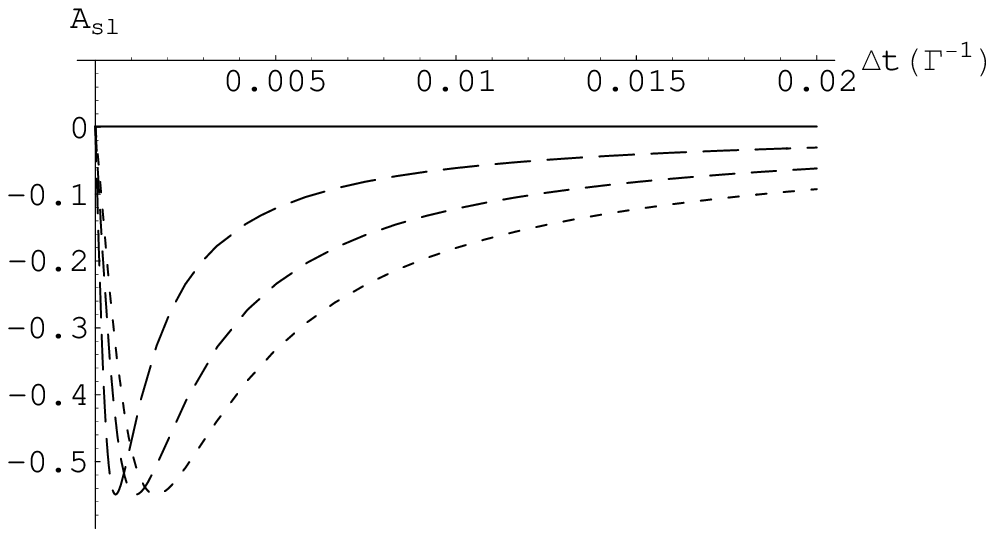}
\caption{\emph{Left picture}: Equal-sign dilepton charge asymmetry $A_{sl}$ for different values of $\omega = |\omega|e^{i\Omega}$, with $\Omega = 0$: $|\omega|=0$ (solid line), $|\omega|=0.0005$ (long-dashed), $|\omega|=0.001$ (medium-dashed), $|\omega|=0.0015$ (short-dashed).  When $\omega\neq0$ a peak of height $A_{sl}(peak) = 0.77 \cos(\Omega)$ appears at $\Delta t(peak)=1.12\, |\omega| \, \frac{1}{\Gamma}$,
$\Gamma = (\Gamma_1 + \Gamma_2)/2$, producing a drastic difference with the $\omega=0$ case, in particular in its time dependence.  Observe that the peak, independently of the value of $|\omega|$, can reach enhancements up to $10^3$ times the value of the asymmetry when $\omega=0$. \emph{Right picture}: as in Left picture, but for $\Omega = 3\pi/2$~\cite{bfactories}.}
\label{asl}
\end{figure}
In B-factories one can look for similar $\omega$-like effects.
Although in this case there is no particularly good channel to lead to enhancement of the sensitivity, as in the $\Phi$-factories, nevertheless one gains in statistics, and hence interesting limits may also be obtained~\cite{bfactories}. The presence of a quantum-gravity induced $\omega$-effect in B systems is associated with a theoretical limitation on flavour tagging, namely the fact that in the absence of such effects the knowledge that one of the two-mesons in a meson factory decays at a given time through a flavour-specific "channel'' determines unambiguously the flavour of the other meson at the same time. This is not true if intrinsic CPT Violation is present.

One of the relevant observables~\cite{bfactories} is given by the CP-violating semi-leptonic decay charge asymmetry (in equal-sign dilepton channel), with the first decay $B \to X\ell^{\pm}$ being time-separated by the second decay
$B \to X'\ell^{\pm}$ by an interval $\Delta t$:
$A_{sl}(\Delta t)  = \frac{I(\ell^+,\ell^+, \Delta t) - I(\ell^-,\ell^-, \Delta t)}{I(\ell^+,\ell^+, \Delta t) + I(\ell^-,\ell^-, \Delta t)} $,
where $I(\Delta t)$ denotes the relevant intensity, integrated over the time of the first decay~\cite{bfactories},
{\scriptsize $I(X \ell^\pm,X' \ell^\pm , \Delta t) = \int_0^\infty \left| \langle X \ell^\pm ,X'\ell^\pm |U(t_1) \otimes U(t_1+\Delta t) |\psi(0)\rangle \right|^2 dt_1 $}, with $U(t)|B_i\rangle =e^{-i(m_i - i\Gamma_i)t}|B_i\rangle~, ~i=1,2 $ the evolution operator for mass-eigenstates states with mass $m_i$ and widths $\Gamma_i$, $i=1,2$. In the absence of $\omega$-effects, the intensity at equal decay times vanishes, $I_{\rm sl}(\ell^{\pm},\ell^{\pm},\Delta t=0) = 0$, whilst in the presence of a complex $\omega=|\omega|e^{i\Omega}$, $I_{\rm sl}(\ell^{\pm},\ell^{\pm},\Delta t=0) \sim |\omega |^2$. In such a case, the function $A_{sl}(\Delta t)$ vs. $\Delta t$ exhibits a peak, whose position depends on $|\omega|$, while the shape of the curve itself depends on the phase $\Omega $ (c.f. figure \ref{asl}).

The analysis of \cite{bfactories}, using the above charge asymmetry method and comparing with currently available experimental data from B-factories on $A_{sl}$:
$A_{sl}^{\rm exp} = 0.0019 \pm 0.0105 $, resulted in the following bounds for the $\omega$-effect:
$ -0.0084 \le {\rm Re}(\omega) \le 0.0100 $ at 95\% C.L. (it is understood that the current experimental
limits give the charge asymmetry as constant, since the relevant analysis has been done in the absence of $\omega$-effects that are responsible for the induced time dependence of this quantity. This has been properly taken into account in \cite{bfactories} when placing bounds).

Before closing we would like to point out that an observation of the $\omega$-effect in both the $\Phi$ and B-factories could also provide an independent test of Lorentz symmetry properties of the intrinsic
CPT Violation, namely whether the effect respects Lorentz symmetry.
This is simply because, although the $\Phi$ particle in neutral Kaon factories is produced at rest,
the corresponding $\Upsilon$ state in B-factories is boosted, and hence there is a frame change betrween the two experiments. If the quantum gravity $\omega$-effect is Lorentz violating, as it may happen in
certain models~\cite{sarkar}, then a difference in value between the two
experiments should be expected, due to frame-dependence, that is dependence on the relative Lorentz factor $\gamma_L$.
\section{Conclusions}
\label{sec:5}
In this work I have discussed a novel phenomenon that may characterise certain quantum gravity models, namely ``intrinsic CPT violation'' as a result of the fact that, due to the associated decoherence of matter propagating in a quantum space-time foam environment, the CPT operator is perturbatively ill-defined: although the anti particle exist, nevertheless the properties of the CPT operator when acting on entangled states of particles lead to modified EPR correlators.
Such modifications imply a set of well-defined observables, which can be measured in current or upcoming facilities, such as $\Phi$ or B-factories. From the non-observation of such modifications at present one places stringent bounds on this type of CPT Violation and the associated quantum gravity decoherence.

The signatures of quantum-gravity induced decoherence in entangled states of mesons are rather unique, and in this sense they constitute ``smoking-gun'' evidence for this type of CPT Violation, if realised in Nature. The other important advantage of such searches is that they are virtually cost free, in the sense that the relevant tests can be performed in facilities that have already been or are to be built for other purposes at no extra cost, apart from minor modifications/adjustments  in the relevant Monte-Carlo programmes to take proper account of these quantum-gravity effects.
\emph{Affaire \'a suivre...}

\section*{Acknowlegements}

I would like to thank the Organisers of the EXA 2008 \& LEAP 2008 joint Conference for the invitation and for providing an excellently organised and thought-stimulating event.
This work is supported in
part by the European Union through the FP6 Marie
Curie Research and Training Network \emph{UniverseNet} (MRTN-CT-2006-035863).

\end{document}